\documentclass[%
 reprint,
 superscriptaddress,
 amsmath,amssymb,
 prl,aps,nofootinbib
]{revtex4-2}

\usepackage[normalem]{ulem}
\usepackage{soul}
\setstcolor{red}

\usepackage{graphicx}
\usepackage{dcolumn}
\usepackage{bm}
\usepackage{braket}
\usepackage{color}
\usepackage[colorlinks=true, allcolors=blue]{hyperref}
\usepackage{orcidlink}

\begin{document}

\title{Dynamical dark energy from an ultralight axion}

\author{Hoang Nhan Luu\,\orcidlink{0000-0001-9483-1099}}
\email{hoang.luu@dipc.org}
\affiliation{Donostia International Physics Center, Basque Country UPV/EHU, San Sebastian, E-48080, Spain}

\author{Yu-Cheng Qiu\,\orcidlink{0000-0002-9008-4564}}
\email{ethanqiu@sjtu.edu.cn}
\affiliation{Tsung-Dao Lee Institute, Shanghai Jiao Tong University, No. 1 Lisuo Road, Pudong New Area, Shanghai, 201210, China}

\author{S.-H. Henry Tye\,\orcidlink{0000-0002-4386-0102}}
\email{sht5@cornell.edu}
\affiliation{Department of Physics, Cornell University, Ithaca, NY 14853, USA}
\affiliation{Department of Physics and Jockey Club Institute for Advanced Study, The Hong Kong University of Science and Technology, Hong Kong S.A.R China}

\date{\today}

\begin{abstract}
Recently the Dark Energy Survey (DES) Collaboration presented evidence that the equation of state $w$ of the dark energy is varying, or $w \simeq -0.948$ if it is constant. In either case, the dark energy cannot be due to a cosmological constant alone. Here we study an ultralight axion (or axion-like particle) with mass $m_\phi \simeq 2 \times 10^{-33}$ eV that has properties that can explain the new $w$ measurement. In particular, $w\ge -1$ and a negative cosmological constant $\Lambda < 0$ is preferred in this model. We also present a simple formula for $w$ for the model to ease data fitting.
 
\end{abstract}

\maketitle
\section{Introduction}

The discovery of the accelerating universe via Type IA supernova measurements implies the existence of dark energy (DE)~\cite{SupernovaSearchTeam:1998fmf, SupernovaCosmologyProject:1998vns}. This is complemented by
one of the greatest successes in cosmology, the precise measurements of the cosmic microwave background (CMB) which support the inflationary-universe paradigm combined with the $\Lambda$-cold-dark-matter ($\Lambda$CDM) model. 
It is known that 70\% of the energy content in our universe is dark energy while the remaining 30\% is matter. 
The simplest interpretation of the dark energy is the introduction of a cosmological constant $\Lambda$ in general relativity, whose equation of state $w_\Lambda = -1$. This is consistent with the CMB measurement of $w=-1.03 \pm 0.03$ by Planck \cite{Planck:2018vyg}. Recently, DES presented evidence that $w$ may be varying, or $w \simeq -0.948$ if it is constant~\cite{DES:2025bxy}. In either case, the DES analysis implies that the dark energy cannot be a cosmological constant alone. With the introduction of an ultralight axion (or axion-like particle), we propose a simple model where $w \ge -1$ is varying, and its property can be tested by improved measurements. 

Ultralight axions are ubiquitous in string theory~\cite{Svrcek:2006yi}. They pick up their exponentially small masses via non-perturbative dynamics (cf.~Ref.~\cite{Hui:2016ltb}), such as the instanton effect. The naturalness of such a light axion in the context of 
Quintessence applied to dark energy has also been extensively studied~\cite{Choi:1999xn,Nomura:2000yk,Kim:2002tq,Hill:2002kq,Rosenfeld:2005mt,Hung:2005ft,Panda:2010uq,Kamionkowski:2014zda,Cicoli:2018kdo,Choi:2021aze,Qiu:2023los,Girmohanta:2023ghm,Wolf:2023uno}. Their presence can resolve some outstanding puzzles in cosmology. The best known case is the fuzzy dark matter (FDM) model, where an axion of mass around $10^{-22}~{\rm eV}$ is the source of the dark matter \cite{Hu:2000ke, Schive:2014dra}. To resolve some issues with the diversity of dwarf galaxies, the introduction of a second axion of mass around $10^{-20}$ eV seems to be necessary~\cite{Luu:2018afg}. Another even lighter axion of mass around $10^{-29}~{\rm eV}$ introduced in the so-called axi-Higgs model~\cite{Fung:2021wbz} helps to explain the $^7$Li puzzle in Big Bang nucleosynthesis (BBN), the Hubble tension and the isotropic cosmic birefringence~\cite{Minami:2020odp}. 

\begin{figure}
    \centering
    \includegraphics[width=7.4cm]{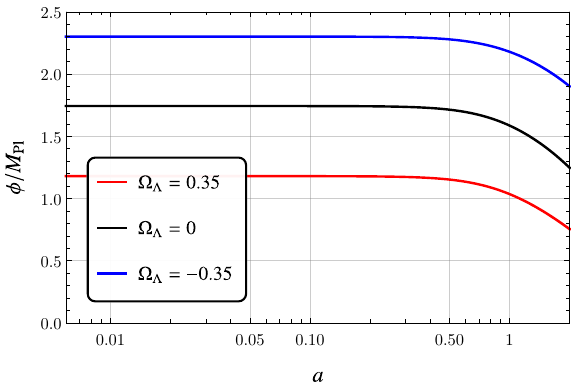}
    \includegraphics[width=7.6cm]{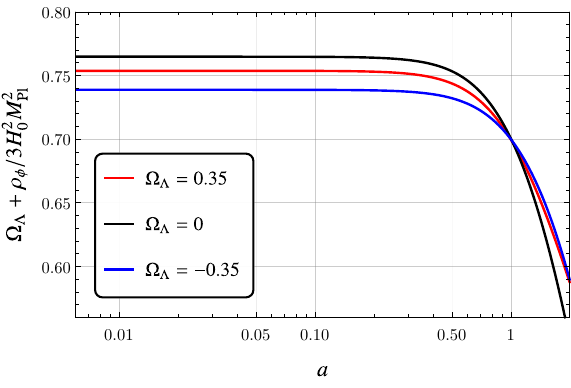}
    \caption{The upper panel is the axion field value evolution against the scale factor, and the lower panel is the effective dark energy density evolution. Here the mass is fixed at $m_\phi =2\times 10^{-33}\,{\rm eV}$, and the fractional energy density of matter is chosen as $\Omega_m =0.3$. Different color corresponds to different choices of $\Omega_\Lambda$.}
    \label{fig:amp_m}
\end{figure}

The introduction of an ultralight axion as dark energy is not new ({\it e.g.}, \cite{Frieman:1995pm, Hlozek:2014lca,Berbig:2024aee,Tada:2024znt,Bhattacharya:2024kxp,Wolf:2024eph}).
Here, we study such an axion $\phi$ with the mass $m_\phi \simeq H_0$ and $f$ taken to be approximately the reduced Planck mass $M_{\rm Pl}$. Most notably, the cosmological constant (vacuum energy) can be negative in the presence of this axion. Let us start with the typical axion potential,  
\begin{equation}
V (\phi) = m_\phi^2f^2\left[1- \cos \left( \dfrac{\phi}{f}\right)\right] \;, 
\label{Eq:axion_potential}
\end{equation}
where $\phi=0$ is at the minimum of the potential ($V(\phi=0)=0$), and $\pi > \phi/f \ge 0$. Suppose the universe starts at $\phi=\phi_i \neq 0$, inflation will lead to $\phi(\bm{x})=\phi_i$ everywhere. (Note that $V(\phi) \simeq m_\phi^2\phi^2/2$ is a good approximation for $\phi_i/f \ll \pi$ that has been usually considered in the literature.) The time evolution of the background field $\phi$ can be derived as
\begin{align}
 \ddot{\phi} + 3H \dot{\phi} + \partial V/\partial\phi =0 \;.
  \label{Eq:ax_eom_t}
 \end{align}
When the Hubble parameter $H \gg m_\phi$, the universe is essentially frozen at the misaligned initial
state $\phi_i$ and $V(\phi_i)$ contribute to the dark energy, {\it i.e.}, the equation of state~(EoS) $w=-1$. As $H \lesssim m_\phi$, $\phi$ starts to roll down along the potential towards $\phi=0$ and deposits the $\phi$ vacuum energy density into matter density (which drops like $a^{-3}$ where $a$ is the scale factor). This behavior is illustrated in Fig.~\ref{fig:amp_m}, where the current dark energy is a combination of a cosmological constant $\Lambda$ and the axion field. That is, $w$ starts to deviate from $w=-1$. In the early stages of the axion rolling, $- 0.9 > w > -1$, and we propose that DES is probing this epoch. 

In this paper, we will confront the proposed axion dark energy (aDE) model with the EoS data of the most recent DES release~\cite{DES:2025bxy}. Our analysis shows that the DES data favors an axion of mass $m_\phi \simeq 2 \times 10^{-33}~{\rm eV}$ and a negative cosmological constant. This result could motivate many string-inspired models where a de Sitter vacuum is not generically expected~\cite{Obied:2018sgi}.

We also find a simple fitting formula for the EoS (with two parameters $w_1$ and $a_1$) that captures the essence of this aDE model~\footnote{The EoS can be generally expressed in a polynomial expansion
\begin{align}
 w(a)= -1+ \sum w_{p}(a-a_1)^{p+1} \;,
\end{align}
where the $p=0$ term is absent due to the smoothness of $w$.
As will be shown later, keeping only the leading $p=1$ term already yields a very good approximation for $a\le 1$,}  
\begin{equation}
    w(a) =
    \begin{cases}
    -1 + w_1 (a-a_1)^2 & 1 \ge a \ge a_1   \\
     -1 & a<a_1
    \end{cases} \; .
    \label{Eq:wfit}
\end{equation}
Roughly speaking, $a_1(m_\phi)$ is a function of the axion mass only while $w_1 (m_\phi, \Lambda)$ also depends on the cosmological constant. For $\Lambda=0$, a good fit yields $w_1 \simeq 0.3$ and $a_1\simeq 0.1$, which corresponds to $m_\phi = 2 \times 10^{-33}~{\rm eV}$. A crucial property of the model is that $w$ never goes below $-1$. If we decrease $m_\phi$, $a_1$ grows so $w$ stays at $w=-1$ for a longer time. Besides, the future of our universe crucially depends on the sign of $\Lambda$. Here $w_1$ decreases for a positive $\Lambda$ (de-Sitter (dS) space) and increases for a (small) negative $\Lambda$ (anti-de Sitter (AdS) space). Since $a_1 \simeq 0.1$, so $w=-1$ during recombination time, the epoch probed by the CMB.
    
\section{Background}

A general cosmological model of an axion, matter and a cosmological constant is written in terms of the first and second Friedmann equations as
\begin{align}
    H^2 &= \dfrac{1}{3M_{\rm Pl}^2}\left(\rho_m + \rho_\phi + \Lambda \right) \;, \label{Eq:fried_t_1} \\
    \dot{H} + H^2 &= -\dfrac{1}{6 M_{\rm Pl}^2}(\rho_m + \rho_\phi + 3p_\phi + \Lambda) \label{Eq:fried_t_2}\;,
\end{align}
where $M_{\rm Pl} = 1 / \sqrt{8\pi G}$. The perfect fluid treatment is applied for all matter species for simplicity. Here $\rho_m$ is the energy density of the total matter (including both baryon and dark matter), which has negligible pressure, $p_m \sim 0$ on large scales. $\rho_\phi$ and $p_\phi$ are the energy density and pressure for the axion field. $\Lambda$ is the vacuum energy density, {\it i.e.}, the cosmological constant, which can be postive, zero, or negative.

The matter density can be expressed as
\begin{equation}
    \rho_m = 3 H_0^2 M_{\rm Pl}^2 \frac{\Omega_m}{a^3} \;, \label{Eq:dm_dens_a}
\end{equation}
where $H_0$ and $\Omega_m$ denote the Hubble parameter and the fractional matter density at the present time ($z = 0$).

The density and pressure of the axion are given by
\begin{align}
    \rho_\phi = \dfrac{1}{2}\dot{\phi}^2 + V(\phi), \quad p_\phi = \dfrac{1}{2}\dot{\phi}^2 - V(\phi) \;.
    \label{eq:ax_dens_press}
\end{align}
Since we are considering an axion field with its mass close to the current Hubble constant, its kinetic energy is much smaller than the potential energy. The axion density $\rho_\phi$ would stay constant like $\Lambda$ for most of the cosmic history.

However, the total dark energy in the aDE model is given by the axion field and the cosmological constant
\begin{align}
\rho_{\rm DE} = \rho_\phi+\Lambda\;, \quad p_{\rm DE}=p_\phi-\Lambda\;, \quad w=\frac{p_{\rm DE}}{\rho_{\rm DE}}\;, \label{Eq:rho_p_w}
\end{align}
where $w=-1$ in the early stage and always $w\gtrsim -1$ at very late times. In the current work, we are particularly interested in the evolution of $w$ as $H$ just drops below $m_\phi$ at the late stage of cosmic evolution.

\begin{figure}
    \centering
    \includegraphics[width=7.6cm]{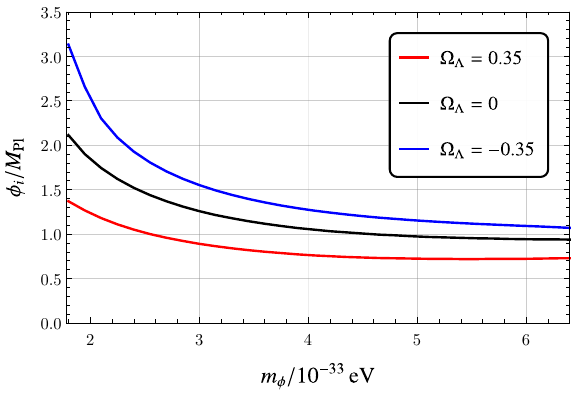}
    \caption{The relation between the axion initial value $\phi_i$ and its mass $m_\phi$ that gives the correct DE relic abundance, $\Omega_\Lambda + \Omega_\phi = 0.7$, where the fractional energy density of the axion field $\Omega_\Lambda$ is defined as $\rho_\phi/3H_0^2 M_{\rm Pl}^2$ at $a=1$. Correspondingly, here we choose $\Omega_m=0.3$.}
    \label{fig:phii_m}
\end{figure}

In practice, it is more convenient to work with scale factor $a$ instead of cosmic time $t$, so we convert all equations above in the following. First is the axion equation of motion~\eqref{Eq:ax_eom_t}, which becomes
\begin{align}
    \phi'' + \left( \dfrac{4}{a} + \dfrac{H'}{H}\right) \phi' + \dfrac{m_\phi^2 f}{a^2 H^2} \sin \left( \dfrac{\phi}{f} \right) = 0\;.\label{Eq:ax_eom_a0}
\end{align}
Here the prime denotes derivative with respect to the scale factor $a$, and we have substituted in the full potential~\eqref{Eq:axion_potential} instead of (only) the mass term.

Combining the two Friedmann equations, one obtains the Hubble friction term, which is
\begin{equation}
    \dfrac{H'}{H} = - \dfrac{3}{2}  \dfrac{H_0^2}{H^2} \dfrac{\Omega_m}{a^4}  - \frac{a \phi'^2}{2M_{\rm Pl}^2} .
\end{equation}
We can then plug this term into Eq.~\eqref{Eq:ax_eom_a0} to get
\begin{align}
    \phi'' + \left( \dfrac{4}{a} - \dfrac{3}{2} \dfrac{\omega_m}{a^4} \dfrac{H_{100}^2}{H^2}  - \frac{a \phi'^2}{2M_{\rm Pl}^2}  \right)\phi' + \dfrac{m_\phi^2f}{a^2 H^2} \sin \left( \dfrac{\phi}{f} \right) = 0\;.\label{Eq:ax_eom_a}
\end{align}
The Hubble parameter is a function of $\phi$ and $\phi'$, which can be obtained through substituting Eq.~\eqref{eq:ax_dens_press} into the first Friedmann equation~\eqref{Eq:fried_t_1},
\begin{multline}
    \dfrac{H^2}{H_{100}^2} = \left( \dfrac{\omega_m}{a^{3}} + \omega_\Lambda + \dfrac{1}{3M_{\rm Pl}^2} \dfrac{m_\phi^2 f^2}{H_{100}^2}\left[ 1 - \cos \left(\dfrac{\phi}{f}\right) \right] \right) \\ \times \left( 1 - \dfrac{a^2 \phi'^2}{6 M_{\rm Pl}^2}  \right)^{-1}. \label{Eq:hubble_a}
\end{multline}
Note that we have defined the dimensionless physical density $\omega_i \equiv \Omega_ih^2$, where $H_0 = H_{100}h$ and $H_{100} = 100~{\rm km}/s/{\rm Mpc}$ ($\simeq 2.1 \times 10^{-33}$ eV) to avoid a circular dependence of $H_0$ when solving Eq.~\eqref{Eq:ax_eom_a} and \eqref{Eq:hubble_a} together. In addition, we define the fractional density of the cosmological constant as $\Omega_\Lambda = \Lambda/3H_0^2 M_{\rm Pl}^2$, and similarly for $\Omega_\phi = \rho_\phi(z=0)/3H_0^2M^2_{\rm pl}$ so that $\Omega_m +\Omega_\Lambda + \Omega_\phi=1$.

Eq.~\eqref{Eq:ax_eom_a} and \eqref{Eq:hubble_a} form a complete set of equations to solve for the axion field $\phi$, {\it including its backreaction on the Hubble flow}. As the axion in our model may account for a large fraction of the matter budget at late times, the influence of the axion dynamics on the background evolution becomes particularly important and should not be neglected. For simplicity we also assume the decay constant $f= M_{\rm Pl}$ for the following analysis.

\section{Analysis}

How to determine the parameters of the aDE model? In general, one shall have four main parameters, including $m_\phi$ and $\phi_i$ for the axion, $\omega_\Lambda$ for the cosmological constant (either positive or negative)
and $\omega_m$ for the matter component. Additionally, we include the baryon density $\omega_b$ as the relevant parameters for early observational data but this parameter is tightly constrained by Big Bang Nucleosynthesis (BBN) measurements, as discussed later.

We constrain the aDE model with the recently released DES data~\cite{DES:2025bxy} via the maximum likelihood method where the log-likelihood function is defined as
\begin{align}
    \log \mathcal{L} \propto \log \mathcal{L}_{\rm BAO} + \log \mathcal{L}_{\rm SN}+ \log \mathcal{L}_\theta + \log \mathcal{L}_t \;. \label{Eq:loglik}
\end{align}
Here the major statistical contributions are from DES observations of Baryon Acoustic Oscillations (BAO)~\cite{DES:2024pwq} and 5-year Supernovae Type IA (SN)~\cite{DES:2024jxu}, represented by the first two terms on the right hand side of \eqref{Eq:loglik}, $\log \mathcal{L}_{\rm BAO}$ and $\log \mathcal{L}_{\rm SN}$, respectively. On the other hand, the remaining terms, $\log \mathcal{L}_\theta$ and $\log \mathcal{L}_t$, are included to enhance the constraining power from derived data of CMB~\cite{Planck:2018vyg} and age of the universe~\cite{Valcin:2020vav}. Let us briefly discuss each of these terms in the following.

Unlike the full DES Y6 BAO likelihood~\cite{DES:2024pwq}, we adopt a Gaussian distribution of the angular scale at an effective redshift for the BAO likelihood. Specifically, we consider the combined constraint on
\begin{align}
    \alpha_\perp \equiv D_M/r_d = 19.51 \pm 0.41 \quad {\rm at} \quad z_{\rm eff} = 0.85, \label{Eq:BAO_constraint}
\end{align}
without assuming any prior on the sound horizon $r_d$. The log-likelihood in this case is simply
\begin{align}
    \log\mathcal{L}_{\rm BAO} \propto -\dfrac{1}{2} \dfrac{( \alpha_\perp - \alpha_{\perp, {\rm obs}} )^2}{\sigma_{\perp, {\rm obs}}^2}, \label{Eq:loglik_BAO}
\end{align}
where $\alpha_{\perp,{\rm obs}}$ and $\sigma_{\perp,{\rm obs}}$ are from \eqref{Eq:BAO_constraint}. We find that applying likelihood method for this compressed BAO observable in place of the full BAO data still yields equally reliable constraints on cosmological parameters. As a proof, we demonstrate this equivalence by reproducing constraints on $w_0, w_a$ and other parameters of the DE model parametrized by Chevallier-Polarski-Linder~(CPL) formula $w = w_0 + w_a(1-a)$ similar to those derived by \cite{DES:2025bxy} in Tab.~\ref{Tab:posterior_bestfit}.

As for SN data, we adopt the full likelihood on the luminosity distance of 1829 SNe Ia from DES-SN5YR~\cite{DES:2024jxu}. Due to the well-known degeneracy between the absolute magnitude $M_0$ and Hubble constant $H_0$, the considered likelihood is obtained after marginalizing over $M_0$, {\it i.e.},
\begin{multline}
    \log \mathcal{L}_{\rm SN} \propto -\dfrac{1}{2}(\bm{\mu}-\bm{\mu}_{\rm obs})^T \bm{C}_{\rm obs}^{-1} (\bm{\mu}-\bm{\mu}_{\rm obs}) \\ + \dfrac{1}{2}\sum_{i,j} (\mu_i - \mu_{{\rm obs}, i}) C^{-1}_{{\rm obs}, ij}/\sum_{i,j} C^{-1}_{{\rm obs}, ij},
\end{multline}
where $\bm{\mu}^T = \{ \mu_1, \mu_2,...\mu_{1829} \}$ is theoretical prediction of the luminosity distance, while $\bm{\mu}_{\rm obs}$ and $\bm{C}_{\rm obs}$ is the observed values and their covariance (including both statistical and systematic uncertainties) measured from SN observations.

We follow Ref.~\cite{DES:2025bxy} to apply the observational constraint from CMB via the angular scale characterizing the position of the first peak in the photon temperature spectrum~\cite{Planck:2018vyg}
\begin{align}
    100\theta_* = 1.04109 \pm 0.00030 \;.
\end{align}
Similarly, we apply the constraint on the age of the universe from data of 38 metal-poor (presumably oldest) clusters~\cite{Valcin:2020vav}, which yields
\begin{align}
    t_{\rm U} = 13.5 \pm 0.5 \; {\rm Gyr}.
\end{align}
These two constraints are necessary to limit the parameter space where early-time and very late-time expansion are not too exotic. As a result, they can together determine $\Omega_m$ and $H_0$ which are not well constrained by BAO and SN data alone. We also assume Gaussian log-likelihood functions for $\theta_*/t_U$, similar to Eq.~\eqref{Eq:loglik_BAO} and denote them $\log\mathcal{L}_\theta/\log\mathcal{L}_t$, respectively.

Note that the real Hubble function in our analysis also includes background contribution from the standard photons and neutrinos ($\sum m_v = 0.06~{\rm eV}$), which do not appear in \eqref{Eq:hubble_a} for simplicity. Another note is that we have used notations such as $D_M, \mu, r_d, \theta_*, t_{\rm U}$ to denote many common cosmological observables without explicit definitions. However, those can be widely found in the literature ({\it e.g.}, see Appendix A of Ref.~\cite{Fung:2021fcj}).

The input parameters $m_\phi, \phi_i, \omega_m$ have uniform priors over the following ranges
\begin{equation} \label{Eq:prior}
\begin{gathered}
    \log_{10}(m_\phi/{\rm eV}) \in [-33, -32], \quad
    \omega_m \in [0, 1] \;, \\ 
    \log_{10}(\phi_i/{\rm GeV}) \in [16, \log_{10}(\pi M_{\rm Pl}/{\rm GeV})]\;.
\end{gathered}    
\end{equation}
The prior of the axion mass as in \eqref{Eq:prior} is motivated to keep the axion DE-like and dynamically relevant before $z = 0$. The axion would be indistinguishable from $\Lambda$ or dark matter when $m_\phi < 10^{-33}~{\rm eV}$ or $m_\phi > 10^{-32}~{\rm eV}$, respectively, at the present time. For $V(\phi)$ \eqref{Eq:axion_potential}, the upper bound of $\phi_i < \pi M_{\rm Pl} \sim 1.5 \times 10^{19}~{\rm GeV}$ is theoretically reasonable given $f = M_{\rm Pl}$.

The physical baryon density $\omega_b$ is another input parameter that can be reliably derived from BBN data. For instance, the latest value from \cite{Schoneberg:2024ifp} reads
\begin{align}
    \omega_b \equiv \Omega_bh^2 = 0.02218 \pm 0.00055 \; . \label{Eq:BBN_data}
\end{align}
As such, we impose a Gaussian prior for $\omega_b$ with mean and variance given by \eqref{Eq:BBN_data}\footnote{The authors of \cite{DES:2025bxy} include the constraint \eqref{Eq:BBN_data} as part of the likelihood, but that is exactly equivalent to treating it as a prior.}.

Since the aDE model may yield multimodal posteriors, we consider two separate cases, one with the anti-de Sitter~(AdS) vacuum, $\Lambda < 0$, and one with the de Sitter~(dS) vacuum, $\Lambda > 0$. The prior of $\omega_\Lambda$ is, therefore, chosen as
\begin{align}
    \omega_\Lambda &\in [0, 1]  &&\text{for an dS vacuum}\;, \\
    \omega_\Lambda &\in [-1, 0] && \text{for an AdS vacuum}\;.
\end{align}

We vary the input parameters of the aDE model via Markov chain Monte Carlo (MCMC) analysis with \texttt{Cobaya}~\cite{Torrado:2020dgo}. A Gelman-Rubin statistic $R-1 < 0.01$ is adopted as the stopping criterion for all MCMC chains. After the run converges, the first $30\%$ steps are discarded as burn-in.

\section{Results}

\begin{figure}[htb]
    \centering
    \includegraphics[scale=0.65]{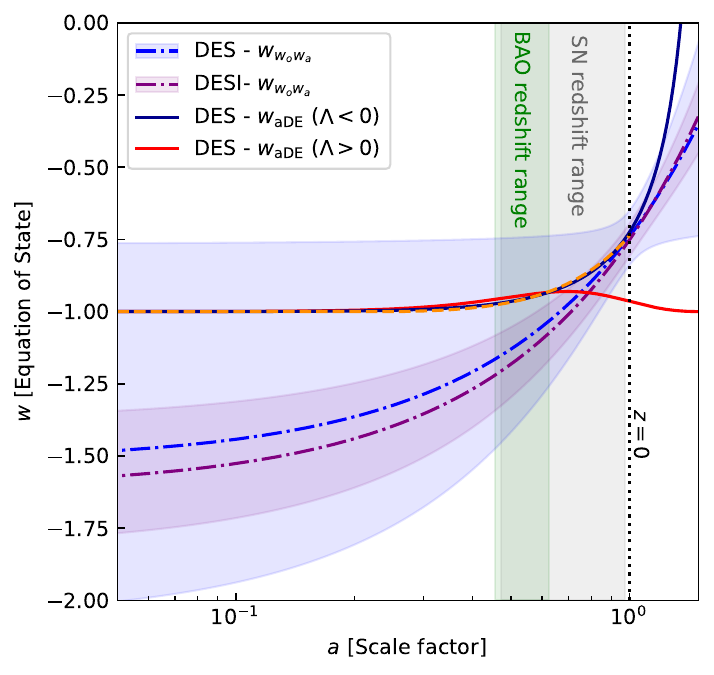}
    \caption{Equation of state $w$ versus the scale factor $a$. The blue dashed-dotted curve and the blue band denote the mean and $1\sigma$ variance of the EoS preferred by DES~\cite{DES:2025bxy}. The recent EoS constraints from DESI~\cite{DESI:2025zgx} is also shown as the purple (dashed-dotted) curve and the purple band for reference. The green and grey bands mark the redshift ranges where DES is probing BAO and SN. The dark blue and red curves are $w$ of the aDE model that has the highest posterior in the AdS and dS case, respectively. The best-fit parameters of each curve can be found in Tab.~\ref{Tab:posterior_bestfit}. The orange (dashed) curve is a phenomenological fit from Eq.~\eqref{Eq:wfit}, which perfectly overlaps with the AdS one up to $a=1$.}
    \label{Fig:ax_de_example}
 \end{figure}

\begin{figure*}[htb]
    \centering
    \includegraphics[scale=0.85]{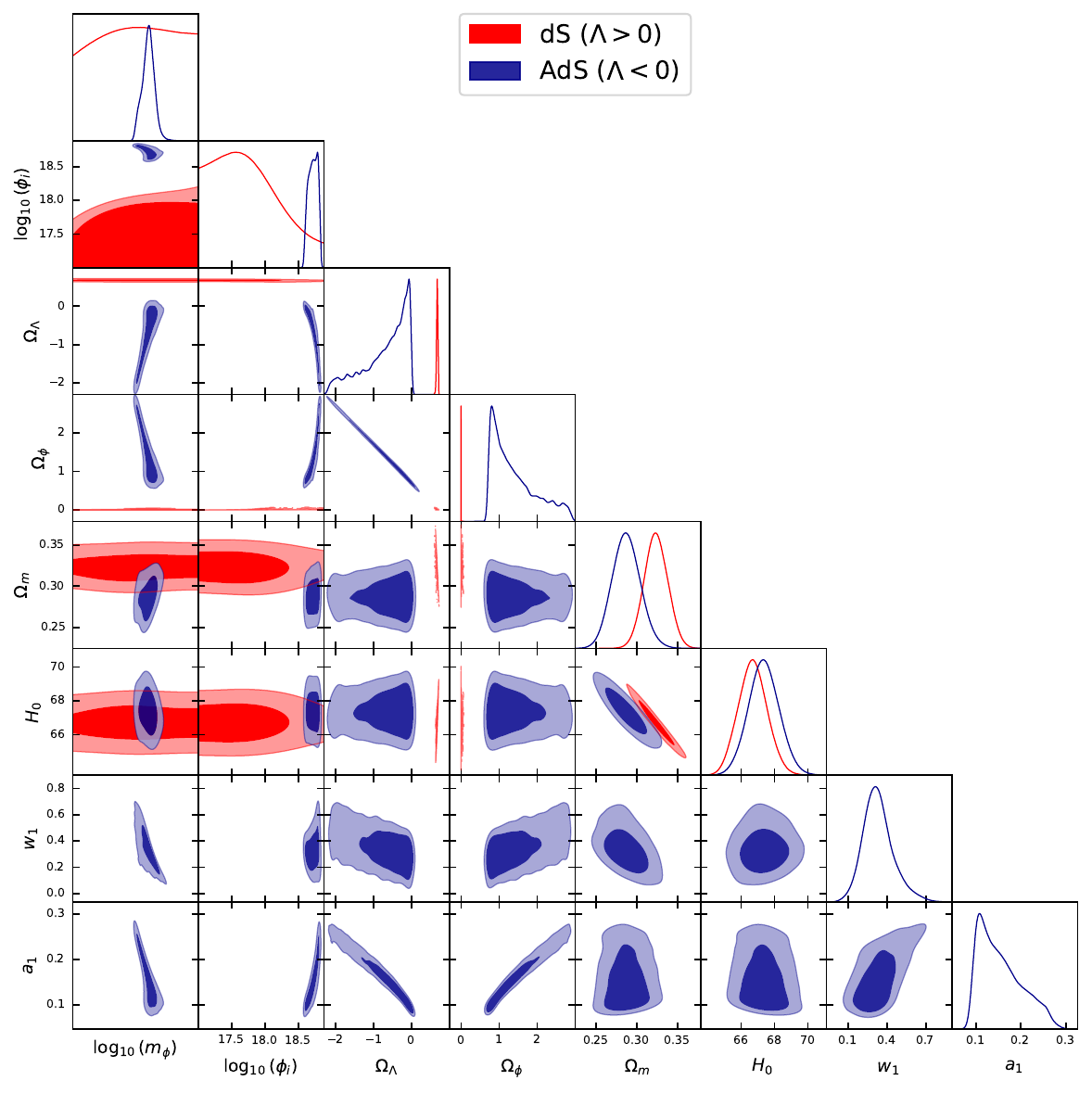}
    \caption{Posterior distributions of parameters in the aDE model. Two contour levels indicate $68\%$ and $95\%$ confidence levels. The red contours indicate the dS model with $\Lambda > 0$, while the blue ones are for the AdS model with $\Lambda < 0$. The units of $m_\phi, \phi_i$ and $H_0$ are implicitly assumed as eV, GeV and ${\rm km}/s/{\rm Mpc}$. The posteriors are computed and created with \texttt{getdist}~\cite{Lewis:2019xzd}.}
    \label{Fig:posterior}
\end{figure*}

In Fig.~\ref{Fig:ax_de_example} the best-fit EoS is shown in each scenario (dS or AdS vacuum). Figure \ref{Fig:posterior} illustrates the corresponding posterior distributions of cosmological parameters. In the following let us highlight several interesting results.

Firstly, if the true vacuum is dS with $\Lambda > 0$, the posterior of $m_\phi$ (in Fig.~\ref{Fig:posterior}) is completely unconstrained while the one of $\phi_i$ shows a weak upper bound. We also see that the distributions of axion and vacuum density cluster around a narrow range of values, {\it i.e.}, $\Omega_\phi \sim 0$ and $\Omega_\Lambda \sim 0.7$, which are almost equivalent to the values preferred in the standard cosmology. In this case, the aDE model with a dS vacumm simply reduces to $\Lambda$CDM. It is, therefore, understandable that $\phi_i$ is constrained to be small and $m_\phi$ becomes irrelevant.

However, the best-fit $w$ for the dS scenario is not constant with respect to the scale factor (see Fig.~\ref{Fig:ax_de_example}) and the best-fit $\Omega_\phi \sim 0.02$ is small but not disappearing (see Tab.~\ref{Tab:posterior_bestfit}). The EoS curve slightly increases and peaks around $z \sim 1$, corresponding to a full oscillation period of the axion. This result proves that DES data indeed disfavor $\Lambda$CDM, hence keeps the aDE (dS) model relevant even with just a small fraction of the axion density.

Secondly, if the true vacuum is AdS with $\Lambda < 0$, we see a clear preference of $m_\phi \sim 2.5 \times 10^{-33}~{\rm eV}$ and $\phi_i \sim 5.2 \times 10^{18}~{\rm GeV}$ (as the mean values, Tab.~\ref{Tab:posterior_bestfit}). Also, the posterior of $\Omega_\phi$ is degenerate with the one of $\Omega_\Lambda$ because the axion density must compensate for a positive cosmological constant to keep the Hubble flow consistent with observational data ($\Omega_\phi + \Omega_\Lambda \sim 0.7$). Unlike the dS case, the axion is now the main driving force for the background evolution. Therefore, we may have more freedom to alter the variation of $w$ (near the present time) from a density drop as the axion enters its first oscillation period. This behavior is the most distinguishing feature of the aDE model with an AdS vacuum compared to the dS one discussed above.

Recall that the axion has already finished one oscillation before $z = 0$ in the best-fit dS case. However, it is still falling down (after being released from the Hubble friction) in the best-fit AdS case. As a result, the DE EoS may increase significantly away from $w = -1$, which in turn yields a better fit with DES data. The total posterior value of the best-fit AdS model is, therefore, higher than the dS one (see Tab.~\ref{Tab:posterior_bestfit}). {\it In other words, a AdS vacuum is more preferred than a dS vacuum for the aDE model.} We emphasize again that the dominance of $|\Omega_\phi|$ with respect to $|\Omega_\Lambda|$ is crucial here because in that case $w = (p_\phi + p_\Lambda)/(\rho_\phi + \rho_\Lambda) \sim p_\phi/\rho_\phi \sim w_\phi$ so that the  EoS of DE is driven by the axion, but not by the cosmological constant. Such a condition best applies when $\Omega_\Lambda \lesssim 0$, {\it i.e.}, when the vacuum is AdS.

Lastly, we notice that the phenomenological formula for $w$ of the aDE model, as proposed in in \eqref{Eq:wfit}, yields a perfect fit to the real axion EoS in the AdS case until $z = 0$ (see Fig.~\ref{Fig:ax_de_example}). On the other hand, it is obvious that this formula would not be a good description for the EoS in the dS case. However, as mentioned above, the aDE model with an AdS vacuum is more generic and more preferred in the light of current data. In that sense, Eq.~\eqref{Eq:wfit} provides a simple parametrization that is ready to be implemented in existing cosmological codes, which could greatly ease data analysis for the aDE model with future experiments.

\begin{table*}[htp]
		\begin{tabular}{c|c|c|c|c|c|c}
			
			Data & \multicolumn{6}{c}{BAO+SN+BBN+$\theta_*$+$t_{\rm U}$}  \\
			\hline
			Model & \multicolumn{2}{c|}{aDE AdS ($\Lambda < 0$)} & \multicolumn{2}{c|}{aDE dS ($\Lambda > 0$)} & \multicolumn{2}{c}{$w_0w_a$CDM} \\
			\hline
			Result & Posterior & Best-fit & Posterior & Best-fit & Posterior & Best-fit \\
			
			\hline\hline
			
			$\log_{10}(m_\phi)$ & $-32.601^{+0.043}_{-0.052}$ & $-32.533$ & unbound & $-32.324$ & - & -  \\
			$\log_{10}(\phi_i)$ & $18.718^{+0.088}_{-0.055}$ & $18.798$ & $< 17.4$ & $17.8$ & - & - \\
			$\omega_\Lambda$ & $> -0.861$ & $-0.731$ & $0.299\pm 0.014$ & $0.297$ & $0.324\pm 0.017$ & $0.329$ \\
			$\omega_m$ & $0.1302\pm 0.0049$ & $0.1293$ & $0.1437^{+0.0030}_{-0.0027}$ & $0.1390$ & $0.133^{+0.010}_{-0.0075}$ & $0.1315$ \\
			$\omega_b$ & $0.02219\pm 0.00055$ & $0.02225$ & $0.02212\pm 0.00054$ & $0.02200$ & $0.02221\pm 0.00055$ & $0.02222$ \\
            $w_0$ & $-0.763\pm 0.072$ & $-0.726$ & $-0.9981\pm 0.0064$ & $-0.9641$ & $-0.730^{+0.086}_{-0.10}$ & $-0.732$ \\
            $w_a$ & - & - & - & - & $-0.78^{+0.78}_{-0.54}$ & $-0.66$ \\
			\hline
			$\Omega_\phi$ & $1.40^{+0.21}_{-0.70}$ & $2.32$ & $0.0035^{-0.0027}_{-0.0037}$ & $0.0130$ & - & - \\
            $\Omega_\Lambda$ & $-0.69^{+0.70}_{-0.20}$ & $-1.61$ & $0.672^{+0.017}_{-0.013}$ & $0.672$ & $0.707 \pm 0.022$ & $0.714$ \\
			$\Omega_m$ & $0.287\pm 0.017$ & $0.284$ & $0.323 \pm 0.014$ & $0.314$ & $0.291\pm 0.022$ & $0.285$ \\
			$\Omega_b$ & $0.0489\pm 0.0016$ & $0.0489$ & $0.0497\pm 0.0016$ & $0.0497$ & $0.0486^{+0.0016}_{-0.0020}$ & $0.0482$ \\
			$H_0$ & $67.36\pm 0.92$ & $67.46$ & $66.70\pm 0.86$ & $66.54$ & $67.6\pm 1.2$ & $67.9$ \\
            $w_1$ & $0.332^{+0.089}_{-0.13}$ & $0.440$ & - & - & - & - \\
            $a_1$ & $0.155^{+0.025}_{-0.063}$ & $0.228$ & - & - & - & - \\
            \hline \hline
            $\log \mathcal{L}_{\rm BAO}$ & $-0.473^{+0.52}_{-0.042}$ & $-0.006$ & $-3.7^{+1.6}_{-1.2}$  & $-2.1$ & $-0.626^{+0.69}_{-0.054}$ & $-0.001$ \\
            $\log \mathcal{L}_{\rm SN}$ & $-819.45^{+0.64}_{-0.073}$ & $-818.807$ & $-821.7^{+1.7}_{-0.65}$ & $-820.35$ & $-819.8^{+1.0}_{-0.28}$ & $-819.03$ \\
            $\log \mathcal{L}_\theta$ & $-0.502^{+0.55}_{-0.039}$ & $-0.009$ & $-0.495^{+0.54}_{-0.043}$ & $-0.021$ & $-0.502^{+0.55}_{-0.038}$ & $-0.002$ \\
            $\log \mathcal{L}_t$ & $-0.206^{+0.044}_{-0.037}$ & $-0.198$ & $-0.186^{+0.042}_{-0.036}$ & $-0.223$ & $-0.186^{+0.15}_{-0.017}$ & $-0.144$ \\
            \hline
            $\log \mathcal{P}$ &  $-821.1^{+1.7}_{-0.56}$ & $-819.09$ & $-826.6^{+1.3}_{-0.56}$ & $-822.77$ & $-821.7^{+1.9}_{-0.73}$ & $-819.24$ \\
		\end{tabular}
	\caption{1D marginalized constraints of cosmological parameters and their best-fit values in the aDE and CPL model, fitted with the combination of BAO+SN+BBN+$\theta_*$+$t_{\rm U}$ similar to Ref.~\cite{DES:2025bxy}. The values quoted are means and $1\sigma~(68\%)$ confidence levels, even for one-sided constraints. Here $w_0$ is not one of the input parameters in the aDE model but it is derived as $w_0 \equiv w(z=0)$. $\log\mathcal{P}$ denotes the final log-posteriord defined as the sum of $\log\mathcal{L}$ in Eq.~\eqref{Eq:loglik} and the BBN prior. An arbitrary normalization factor is adopted for all likelihood, prior, posterior quantities. Note that some best-fit parameters of the aDE AdS model do not lie within 1$\sigma$ of the corresponding mean values because these distribution are highly skewed towards one side of their prior ranges (see Fig.~\ref{Fig:posterior}). The units of $m_\phi, \phi_i$ and $H_0$ are implicitly assumed as eV, GeV and ${\rm km}/s/{\rm Mpc}$.}
	\label{Tab:posterior_bestfit}
\end{table*}



\section{Discussion}

In summary, as shown in Tab.~\ref{Tab:posterior_bestfit},
in demanding $\Lambda > 0$, the best-fit EoS is consistent with $\Omega_{\Lambda}=0.7$, implying that the axion field is not needed and $w = -1$ is acceptable. However, the recent DES and DESI measurement suggest $w \ne -1$ at up to $4 \sigma$ level, so we are strongly motivated to explore the aDE model on the $\Lambda < 0$ region, which has never been constrained by observational data. Interestingly, this regime provides a fit to the DES data that is comparable in quality to that of the $w_0w_a$CDM model with the best-fit parameters given by
\begin{equation}
    \begin{aligned}
    \begin{gathered}
        m_a = 2.93 \times 10^{-33}~{\rm eV}, \quad \phi_i = 6.28 \times 10^{18}~{\rm GeV}, \\
    \Omega_\phi = 2.32, \quad \Omega_\Lambda = -1.61 \quad \rightarrow \quad \Omega_{\rm DE} = 0.71, \\
    \Omega_m = 0.284, \quad H_0 = 67.46~{\rm km}/s/{\rm Mpc}, \\
    w_1 = 0.44, \quad a_1 = 0.23.
    \end{gathered}
    \end{aligned}
\end{equation}

That the vacuum energy density of the universe might drop over time is natural in the stringy axiverse. Thanks to the misalignment mechanism, the repeated drops of the vacuum energy density is expected in any model with multiple light axions, as every axion will lead to a drop in the vacuum energy density, as described by Eq.~\eqref{Eq:ax_eom_t}. The question is what issues can be resolved with the introduction of a specific axion.

If the EoS indeed goes below $w < -1$ as allowed by the DES and DESI data, this poses two related challenges: (1) an exotic theoretical proposal is needed to realize this; (2) $w=-1$ is observed by CMB measurements at $a \sim 10^{-4}$, so $w < -1$ at $a \sim 0.1$ requires $w$ to drop before rising again. A new mechanism is needed to explain this behavior. By comparison, the aDE model is simple and well motivated within the conventional quantum field theory framework, strongly motivated by our understanding of string theory.

The aDE model surprisingly favor $\Lambda < 0$ in light of the most recent DES data. If that is the case, the universe will eventually end in an AdS space. Many string theorists believe that the ground state of our universe is supersymmetric, which must have a negative vacuum energy density. With the introduction of an ultralight axion, which is quite acceptable in string theory, we now have a simple way to realize this belief. In fact, our model distinguishes itself from the so-called ``thawing dark energy''~\cite{Caldwell:2005tm, Lodha:2025qbg} by this possibility of a negative cosmological constant. However, as pointed out by Coleman \& de Lucia~\cite{Coleman:1980aw}, the universe with a negative $\Lambda$ will end in a big crunch. To this day, we do not know how this big crunch can be avoided if the universe is indeed supersymmetric.

Lastly, we should also mention the existing Hubble tension. Obviously, the aDE model considered here cannot fit both $w$ preferred by DES and $H_0$ preferred by SH0ES~\cite{Riess:2021jrx} at the same time. There is an idea that $H_0$ and $w_0$ must be anti-correlated in any quintessence models that satisfy the most recent DES/DESI and other observations~\cite{Lee:2022cyh}. This means any data-fitting attempt that leads to $w_0 > -1$ would also worsen the Hubble tension. In this study, we found that this effect manifests in the dS model but not in the AdS one (see $H_0$ in Tab.~\ref{Tab:posterior_bestfit}). On the other hand, the recently proposed axi-Higgs model~\cite{Fung:2021wbz}, with an ultralight axion of mass $m_\phi \simeq 10^{-29}~{\rm eV}$, can lift the electron mass during recombination time, hence ameliorates the Hubble tension. We look forward to a future analysis where both $H_0$ and $w$ tension will be simultaneously resolved.

\section*{Acknowledgment}
The work of Y.-C.~Qiu is supported by the K.~C.~Wong Educational Foundation.

\bibliographystyle{utphys}
\bibliography{reference}

\end{document}